\documentclass[twocolumn,aps,prb,floatfix,preprintnumbers,
showpacs,superscriptaddress]{revtex4}

\makeatletter
\usepackage{graphicx}
\usepackage{amsmath}
\usepackage{amssymb}
\usepackage{amscd}
\usepackage{bm}
\makeatother

\begin{document}

\title{Spin, charge, and orbital correlations \\
in the one-dimensional $t_{\rm 2g}$-orbital Hubbard model}

\author{J. C. Xavier}
\affiliation{Condensed Matter Sciences Division,
Oak Ridge National Laboratory, Oak Ridge, TN 37831 and
\\ Department of Physics,
The University of Tennessee, Knoxville, TN 37996}
\affiliation{Universidade Estadual Paulista,
CP 713, 17015-970 Bauru, SP, Brazil}

\author{H. Onishi}
\affiliation{Advanced Science Research Center,
Japan Atomic Energy Research Institute,
Tokai, Ibaraki 319-1195, Japan}

\author{T. Hotta}
\affiliation{Advanced Science Research Center,
Japan Atomic Energy Research Institute,
Tokai, Ibaraki 319-1195, Japan}

\author{E. Dagotto}
\affiliation{Condensed Matter Sciences Division,
Oak Ridge National Laboratory, Oak Ridge, TN 37831 and
\\ Department of Physics,
The University of Tennessee, Knoxville, TN 37996}

\date{July 20, 2005}

\begin{abstract}
We present the zero-temperature phase diagram of the one-dimensional
$t_{\rm 2g}$-orbital Hubbard model, obtained using
the density-matrix renormalization group and Lanczos techniques.
Emphasis is given to the case for the electron density $n$=5
corresponding to five electrons per site,
of relevance for some Co-based compounds.
However, several other cases for electron densities
between $n$=3 and 6 are also studied.
At $n$=5, our results indicate a first-order transition between
a paramagnetic (PM) insulator phase and a fully-polarized
ferromagnetic (FM) state by tuning the Hund's coupling.
The results also suggest a transition from the $n$=5 PM insulator phase
to a metallic regime by changing the electron density,
either via hole or electron doping.
The behavior of the spin, charge, and orbital correlation functions
in the FM and PM states are also described in the text and discussed.
The robustness of these two states varying parameters suggests that
they may be of relevance in more realistic higher dimensional systems
as well.
\end{abstract}



\maketitle


\section{Introduction}

The study of the exotic properties of cobalt oxides is an area of
investigations that is currently attracting considerable attention
in the research field of condensed matter physics.
Among the main reasons for this wide effort,
the recent discovery of superconductivity
in layered two-dimensional triangular lattices of Co atoms
with the composition Na$_{\rm x}$CoO$_2$ has certainly triggered
a rapid increase of research activities on cobalt oxides.
This material becomes superconducting after H$_2$O is
intercalated,\cite{cobalt-SC}
opening an exciting area of investigations.

The experimentally unveiled phase diagram of this compound
varying the Na composition has revealed the existence of
several other competing tendencies:
Charge ordered as well as magnetic states are stabilized,
in addition to superconductivity.\cite{cobalt-cava}
In the related compound $\rm (Ca_2 Co O_3)(CoO_2)$,
a spin incommensurate spin-density-wave has been recently
reported.\cite{cobalt-IC}
The existence of such a rich phase diagram is a characteristic
of strongly correlated electron systems,
where complex behavior typically emerges due to the presence
of competing states that have similar energies
but vastly different transport and magnetic properties.\cite{complexity}

Additional motivation for the study of Co-oxides arises from
recent experimental studies of hole-doped cobaltites
in the perovskite form such as $\rm La_{1-x}Sr_{x}CoO_3$,
where clear tendencies toward phase separation between
ferromagnetic (FM) metallic and paramagnetic (PM) insulating
regions have been found.\cite{cobalt-PS}
This establishes an intriguing qualitative connection
between Co-oxides and the famous manganites that exhibit
the colossal magnetoresistance,
effect widely believed to originate in an analogous mixed-phase
tendency exhibited by Mn-oxides.\cite{book}
In fact, a large magnetoresistance in some cobaltites has
also been observed, and its origin appears related with
phase competition.\cite{raveau}
As a consequence, establishing the dominant ground-state
tendencies of simple models for cobaltites is important
to envision the possible phase mixtures
that may lead to exotic behavior.

As a third motivation for the study of cobaltites,
it is known that some Co-based compounds have interesting
thermoelectric properties.
In particular, a huge thermoelectric power has been recently
discovered in NaCo$_2$O$_4$ by Terasaki {\it et al.},\cite{terasaki}
opening another area of investigations,
with a focus on thermoelectric materials
mainly with the purpose of industrial applications.

For all these reasons, the theoretical study of models for Co-oxides
is timely and needed in order to guide further experimental developments.
Ab-initio calculations have already provided important information
in this context,\cite{singh}
and the inclusion of many-body effects is the natural next step.
Previous theoretical studies of Co-based systems including
Coulombic repulsion have mainly focused on triangular lattices.
In this context, recent Monte Carlo investigations unveiled
the presence of magnetic correlations.\cite{maekawa}
Fluctuation-exchange approximations also revealed tendencies
toward ferromagnetism and possible triplet-pairing instabilities
in a multiorbital model.\cite{ogata}
Several approximate studies of $t$-$J$~\cite{single-band-t-J}
and single-band Hubbard models~\cite{single-band-Hubbard}
have also been presented.

To understand the behavior of complex oxides, it is of particular
importance the analysis of the many possible tendencies
in the ground state, namely the study of the various competing
states stabilized as electron density and coupling are modified.
Unfortunately, this task is difficult due to a lack of reliable
unbiased analytical techniques.
For this reason, the first effort toward a detailed numerical
analysis of models for cobaltites is presented here.
Instead of directly emphasizing the triangular lattice with
approximate techniques or exactly studying small systems,
we have preferred to perform a systematic study of
a one-dimensional multiorbital Hamiltonian,
exploring in detail the coupling and electron density
parameter space, and using computationally exact techniques.
This level of accuracy is achieved through the use of reliable methods
such as the density-matrix renormalization group (DMRG)~\cite{white}
and the Lanczos technique.\cite{review}
We envision this effort as a first step toward
a systematic computational analysis of more complicated
quasi-two-dimensional triangular-lattice systems.

The paper is organized as follows.
In Sec.~II, the multiorbital model is introduced and
many-body computational techniques used here are briefly discussed.
In Sec.~III, the main results are presented.
These results are organized based on the observable studied:
First, the $n$=5 phase diagram is discussed,
where $n$ denotes the number of electron per site.
Then, the spin correlations are presented at several values of $n$.
This is followed by the charge and orbital correlations.
Finally, conclusions are presented in Sec.~IV.
The main result of the paper is the clear dominance of
two rather different ground states: (1)
a FM state and (2) a PM state with short-range correlations.
Both are very robust varying couplings and densities.
Their higher-dimensional versions may be of relevance for
present and future Co-oxide experiments.


\section{Model and Technique}

In the investigation reported in this manuscript, we consider
a three-orbital Hubbard model, defined on a one-dimensional chain
along the $x$-axis with $L$ sites.
The three orbitals represent the $t_{\rm 2g}$ orbitals of relevance
for cobaltites.
The model is given by
\begin{eqnarray}
 H &=&
 -\sum_{j,\gamma,\gamma',\sigma}t_{\gamma,\gamma'}
  \left(d_{j,\gamma\sigma}^{\dagger}
        d_{j+1,\gamma'\sigma}^{\phantom{\dagger}}+
       \mathrm{H.}\,\mathrm{c.}\right)
 \nonumber\\
 &&+U\sum_{j,\gamma}\mathbf{\rho}_{j,\gamma\uparrow}
    \mathbf{\rho}_{j,\gamma\downarrow}
   +\frac{U'}{2}\sum_{j,\sigma,\sigma',\gamma\ne\gamma'}
    \mathbf{\rho}_{j,\gamma\sigma}\mathbf{\rho}_{j,\gamma'\sigma'}
 \nonumber\\
 &&+\frac{J}{2}\sum_{j,\sigma,\sigma',\gamma\ne\gamma'}
    d_{j,\gamma\sigma}^{\dagger}d_{j,\gamma'\sigma'}^{\dagger}
    d_{j,\gamma\sigma'}d_{j,\gamma'\sigma}
 \nonumber\\
 &&+\frac{J'}{2}\sum_{j,\sigma\ne\sigma',\gamma\ne\gamma'}
    d_{j,\gamma\sigma}^{\dagger}d_{j,\gamma\sigma'}^{\dagger}
    d_{j,\gamma'\sigma'}d_{j,\gamma'\sigma},
\end{eqnarray}
where the index $j$ denotes the site of the chain,
$\gamma$ indicates the orbitals $xy$, $yz$, and $zx$,
and $\sigma$ is the spin projection along the $z$-axis.
The rest of the notation is standard.
The hopping amplitudes are $t_{xy,xy}$=$t_{zx,zx}$=$t$=$1$,
and zero for the other cases.
These simple values for the hopping amplitudes can be easily
derived from the overlap of $d_{xy}$, $d_{yz}$, and $d_{zx}$
orbitals between nearest-neighbor sites along the $x$-axis.
The interaction parameters $U$, $U'$, $J$, and $J'$ are the standard
ones for multiorbital Hamiltonians, and a detailed description
can be found in Ref.~\onlinecite{book}.
These couplings are not independent, but they satisfy the well-known
relations $J'$=$J$ and $U$=$U'$+$2J$, due to the reality of
the wave function and the rotational symmetry in the orbital space.

We investigate the model described above mainly using
the DMRG technique with open boundary conditions.\cite{white}
The finite-size algorithm is employed for sizes up to $L$=$48$,
keeping up to $m$=$350$ states per block.
The truncation errors are kept around $10^{-5}$ or smaller.
The center blocks in our DMRG procedure are composed of 64 states
due to the three orbitals.
Note that, for instance, the $t$-$J$ model has only 3 states
in these center blocks.
As a consequence, keeping $m$=$350$ states per block
in the $t_{\rm 2g}$-orbital Hubbard model is analogous to
keeping $m$$\sim$$7000$ states per block in the $t$-$J$ model.

Although in related investigations specific values of $U$ and $J$
for the triangular Co-oxides were discussed,\cite{ogata}
here we prefer to vary independently these couplings analyzing
the possible ground states that are stabilized by this procedure.
In fact, the ratio $U/J$ may change among the many interesting Co-oxides,
and in addition, it is important to classify the states that
could be stabilized by the proper isovalent chemical doping,
external fields, or perturbations.


\section{Results}


\subsection{Phase diagram for density $n$=5}

\begin{figure}[t]
\begin{center}
\includegraphics[width=0.8\linewidth]{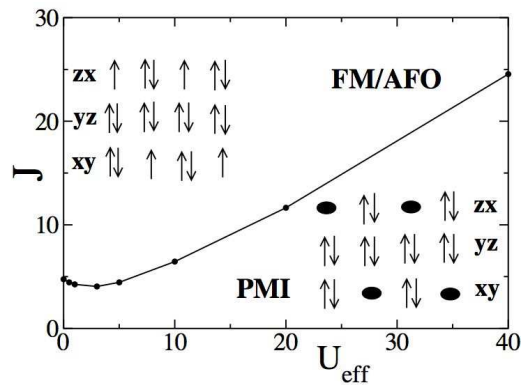}
\end{center}
\caption{\label{fig1}
Ground-state phase diagram for the one-dimensional three-orbital Hubbard
model, using a 6-site chain and working at electron density $n$=5.
FM and PMI denote the regions with ferromagnetism and paramagnetism
(insulator), respectively.
We also present a schematic picture of the electron configurations.
AFO indicates the staggered population of orbitals in the FM state.
The reader should consult the text for more details,
as well as the next figure.
}
\end{figure}

In Fig.~1, the ground-state phase diagram $J$ versus
$U_{\rm eff}$=$U'$$-$$J$ is presented.
For large $J$, a fully polarized FM phase is obtained,
while for small $J$, a PM regime is found.
This PM phase is insulating, as shown below.
Note that the phase diagram is obtained by comparing
the energies for different sectors of the $z$-projection
of the total spin, $S_{\rm total}^{z}$, mainly using a system of size $L$=6.
Other values of $L$ are also studied, and it is observed that
for $L$=4, 6, 8, and 10, in the PM regime the ground state has
total spin 0, 1, 0, and 1, respectively,
for a large set of couplings investigated.
As a consequence, it is reasonable to assume that
the transition line separates states with the minimum and maximum
total spin, without intermediate partially polarized regimes.
The phase diagram we have found has similarities with that already
reported by two of the authors at density $n$=$4$,\cite{onishihotta}
in the context of spin-1 chains.

As described later,
our results for the spin-spin, charge-charge, and orbital
correlations suggest, roughly, an electron distribution
at short distances, as schematically presented in Fig.~1.
The electron configuration in the FM phase is quite simple:
5 electrons per site, with a polarized net spin 1/2
and antiferro-orbital (AFO) correlations.
The existence of FM correlations is a direct consequence
of the multiorbital nature of the model and
the robust value of $J$ in the FM regime.

\begin{figure}[t]
\begin{center}
\includegraphics[width=0.5\linewidth]{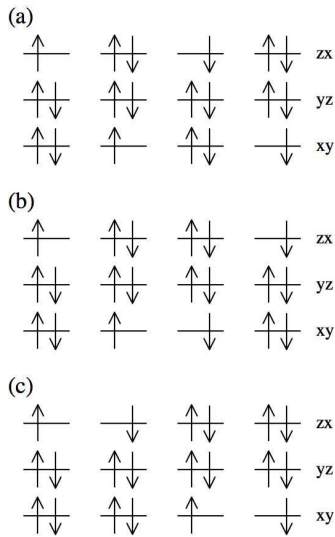}
\end{center}
\caption{\label{fig2}
States with the largest weight in the ground state of
a 4-site chain solved exactly.
Note that each state has 8-fold degeneracy.
At $J$=0, these three states have the same weight.
On the other hand, for nonzero $J$, the state (a)
(and its 8-fold degenerate states) has the largest weight,
with a spin (orbital) structure factor peaked at $\pi/2$ ($\pi$).
The states (b) and (c) (each one also with degeneracy 8) have
the second- and third-largest weights, respectively, for nonzero $J$.
}
\end{figure}

On the other hand, a more complex electron configuration
emerges in the PM phase.
The meaning of the full circles in the inset of Fig.~1
for the PM phase is to denote either a spin up or a spin down.
Note, however, that quantum fluctuations are strong and
the configuration shown in Fig.~1 is just a guidance.
To obtain insight into the ground-state wave function,
it is useful to consider the case of a four-site chain,
where results can be obtained exactly by using the Lanczos method.
In the strong-coupling limit $U_{\rm eff} \gg J \gg 1$
(or, more precisely, $1/(U'-J) \ll 1$),
it is found that the most important portion of
the ground-state wave function is expressed as
\begin{eqnarray}
 |\psi\rangle&=&\frac{1}{\sqrt{24}}\sum_{\rm P}(-1)^{n_{\rm P}}
 \nonumber\\
 && \times
 \left(\begin{array}{c}
 \uparrow\downarrow\\
 \uparrow\downarrow\\
 \downarrow
 \end{array}\right)
 \otimes
 \left(\begin{array}{c}
 \downarrow\\
 \uparrow\downarrow\\
 \uparrow\downarrow
 \end{array}\right)
 \otimes
 \left(\begin{array}{c}
 \uparrow\downarrow\\
 \uparrow\downarrow\\
 \uparrow
 \end{array}\right)
 \otimes
 \left(\begin{array}{c}
 \uparrow\\
 \uparrow\downarrow\\
 \uparrow\downarrow
 \end{array}\right),
\end{eqnarray}
where the sum is taken over the permutation of the four spinors and
$n_{\rm P}$ is the number of permutation we have to perform to recover
the original configuration.
Namely, the electron configuration presented 
in the PM phase of Fig.~1 should be regarded
as the equivalent of the 4 spinors contained in $|\psi\rangle$.
Note that this is not a rigid configuration,
but all permutations are equally important at small $J$.
In particular, all the 24 states have the same weight in the ground state
at $J$=0, while at finite $J$, the 24 states are split into three classes
with 8 states for each, as shown in Fig.~2.
Note that each of these classes
lead to a distinct peak position in the spin and orbital
structure factors.
When the peak positions in these channels are denoted
by $q_{\rm spin}$ and $q_{\rm orbital}$,
the class (a) has $q_{\rm spin}$=$\pi/2$ and $q_{\rm orbital}$=$\pi$,
class (b) $q_{\rm spin}$=$\pi/2$ and $q_{\rm orbital}$=$\pi/2$,
and class (c) $q_{\rm spin}$=$\pi$ and $q_{\rm orbital}$=$\pi/2$.
The $yz$ orbital is fully occupied
due to the one dimensionality of the system that prevents
the movement of electrons in this orbital due to a vanishing 
hopping.
For the active $xy$ and $zx$ orbitals, the electrons are not distributed
in a rigid charge-ordered pattern, but instead the density is
to an excellent approximation equal to 1.5 at every site.
Note that a similar representation of the ground-state wave function
for four sites has been found for the $SU(4)$ spin-orbital
model.\cite{zhangsu4}
As discussed in more detail below, these two models are
related to each other.


\subsection{Spin correlations at several densities}

To understand more quantitatively the magnetic order present in the PM
phase, it is useful to measure the spin-spin correlation function,
defined as
\begin{equation}
  C_{\rm spin}(l)=\frac{1}{M}\sum_{|i-j|=l}
  \left\langle S_{i}^{z}S_{j}^{z}\right\rangle,
\end{equation}
where $S_{i}^{z}$=%
$\sum_{\gamma}(\rho_{i\gamma\uparrow}-\rho_{i\gamma\downarrow})/2$
is the $z$-projection of the total spin at each site and
$M$ is the number of site pairs $(i,j)$ satisfying $l$=$|i-j|$.
We average over all pairs of sites separated by distance $l$,
in order to minimize boundary effects.
In Fig.~3(a), $C_{\rm spin}(l)$ is shown for the PM phase.
It is found that the numerical data of $C_{\rm spin}(l)$
are well reproduced by the function
\begin{equation}
  \label{Eq:fit}
  {\tilde C}_{\rm spin}(j)=
  \frac{a}{j^{2}}+b\frac{\cos(\frac{\pi}{2}j)}{j^{3/2}},
\end{equation}
as shown by the dashed curve.
Note that $a$ and $b$ are appropriate fitting parameters.
The result indicates that
the spin-spin correlation function has a four-site periodicity
and decays as a power law with critical exponent 3/2.
In the inset of Fig.~3(a), we also present the Fourier
transform of the spin-spin correlation function,
\begin{equation}
  S\left(q\right)=
  \frac{1}{L}\sum_{j,k}e^{iq\left(j-k\right)}
  \left\langle S_{j}^{z}S_{k}^{z}\right\rangle,
\end{equation}
for $L$=$16$ and $L$=$48$.
As observed in this figure, finite-size effects appear to
be very small.
Here we clearly find a peak in $S(q)$ at $q$=$\pi/2$,
corresponding to the four-site periodicity of
the spin-spin correlation function.

\begin{figure}
\begin{center}
\includegraphics[width=0.75\linewidth]{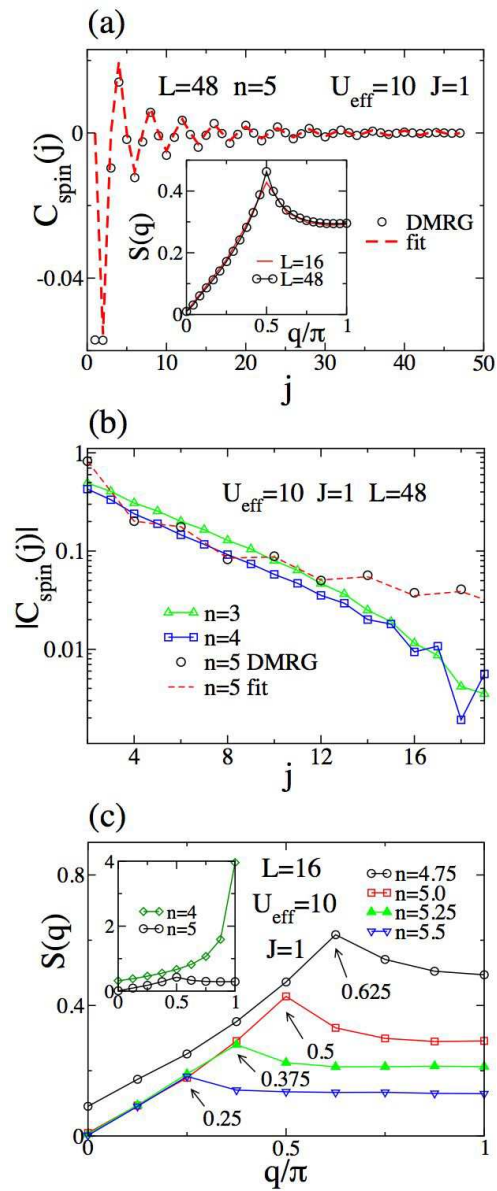}
\end{center}
\caption{\label{fig3}
(a) The spin-spin correlation function $C_{\rm spin}(j)$ vs. $j$
for $L$=$48$ and density $n$=$5$.
The dashed line indicates a fit using Eq.~(\ref{Eq:fit}).
The inset shows the spin structure factor $S\left(q\right)$ for
$L$=$16$ and $L$=$48$.
(b) The linear-log plot of the module of the spin-spin correlation
$|C_{\rm spin}(l)|$ corresponding to densities $n$=$3$, 4, and 5
with $L$=$48$, as well as the fit used in (a).
For details, see the main text.
(c) Spin structure factor $S\left(q\right)$ for several densities $n$,
and using $L$=$16$.
The arrows indicate the peak positions.
In all plots $U_{\rm eff}$=$10$ and $J$=$1$, as indicated.
Inset shows $S(q)$ for $n$=$4$ and 5.
}
\end{figure}

Let us here discuss the physical meaning of the four-site periodicity.
Since the $yz$ orbital is fully occupied in our studies,
the $t_{\rm 2g}$-orbital Hubbard model can be regarded as
a two-orbital Hubbard model composed only of $xy$ and $zx$ orbitals.
Note that for this two orbital model,
the hopping amplitudes are symmetric and there is no off-diagonal elements.
Moreover, when $J$=$0$, in the two orbital model,
there exists an extra $SU(4)$ symmetry involving both spin and orbital
degrees of freedom.
In such a case, the effective Hamiltonian in the strong-coupling limit
is given by the $SU(4)$ spin-orbital model,
which has been investigated intensively
in recent years.\cite{Troyer1,shibatasu4b,afflecksu4}
Concerning a less symmetric case than $SU(4)$,
the effect of $J$ has also been discussed,\cite{shibatasu4,boulatsu4}
where anisotropic exchange interactions arise in the orbital part.
Note that the spin-spin correlation function, presented in Fig.~3(a),
is found to have a four-site periodicity and decay as a power law
with critical exponent 3/2.\cite{comment-3/2}
These results are consistent with previous
analytical work~\cite{afflecksu4b} and
numerical analysis~\cite{Troyer1,shibatasu4}
for the $SU(4)$ spin-orbital model.

Here we stress that the spin-spin correlation functions for $n$=$5$
clearly present distinct behavior from the results already reported
at $n$=$4$,\cite{onishihotta} where an exponential decay has been
observed, as depicted as a linear-log plot in Fig.~3(b).
The result indicates a $gapless$ spin-excitation spectrum for $n$=$5$,
with power-law decaying correlations, in contrast to a gapfull behavior
for $n$=$4$.
Note that, for better comparison, we have normalized $C_{\rm spin}$
in such a way that the correlations are the same at distance one.
We have eliminated the odd sites for $n$=$5$, since the results there
are close to zero (see Fig.~3(a)).
Note also that working with $m$=$350$, it is difficult to reach
good accuracy for the correlations at large distances,
since they are very small. For this reason, in our results we present
only the first 19 sites.\cite{comment-gap}

In Fig.~3(b), we also show the spin-spin correlation function for $n$=3,
reported here for the first time to our knowledge.
In the case of $n$=3, it is naively expected that the local spin $S$=3/2
is formed at each site.
By analogy with the half-odd-integer-spin antiferromagnetic Heisenberg chains,
we expect the power-law decay of the spin-spin correlation function and
a gapless spin-excitation spectrum as well.\cite{Haldane-1983}
On the contrary, we can observe in Fig.~3(b) that the spin-spin correlation
function shows an exponential decay similar to the case
for the integer-spin chains.
To understand this peculiar behavior, it is necessary to take into account
the effect of $t_{\rm 2g}$ orbitals.
As mentioned above, electrons in the $yz$ orbital cannot hop,
while electrons in the $xy$ and $zx$ orbitals move to adjacent sites
with the same amplitude.
Then, it is expected that only electrons in the $xy$ and $zx$ orbitals
contribute to the exchange interaction, and the $n$=3 system is
regarded as an effective $S$=1 chain, leading to the exponential decaying
spin-spin correlation function.

In Fig.~3(c), $S(q)$ is shown for several densities.
Since the finite-size effects seem to be small, we consider $L$=$16$.
As observed in these studies, the results suggest that the peak position
changes linearly with the electron density as
$q$=$(6-n)\pi/2$ (mod $\pi$).
Note that this peak is clearly robust for $n$=$4$, as shown
in the inset of Fig.~3(c), and substantially decreases its intensity
by increasing the density $n$.

It is important to remark that the inset of Fig.~3(a) is very similar
to the results found by Ogata and Shiba in their study of
the one-dimensional Hubbard model at quarter-filling and $U$=$\infty$
(see Fig.~9 of Ref.~\onlinecite{ogata-shiba}).
Clearly, in the model studied in this paper, the electrons
in the two bands with a nonzero hopping behave like
one-band models with a strong on-site repulsion,
at least from the perspective of the spin correlations.
Note, however, that these two one-band models are connected
via the Coulombic repulsion which, as discussed below,
will open a gap in the spectrum of charge excitations.


\subsection{Charge correlations at several densities}

\begin{figure}
\begin{center}
\includegraphics[width=0.8\linewidth]{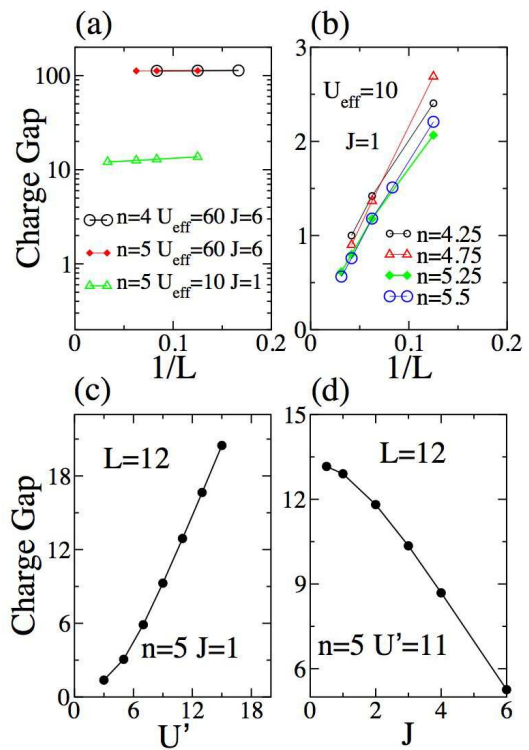}
\end{center}
\caption{\label{fig4}
(a) The charge gap $\Delta$ vs. $1/L$ at particular values of
$U_{\rm eff}$ and $J$, and densities $n$=$4$ and $n$=$5$.
(b) Same as (a) but for non-integer densities, and
$U_{\rm eff}$=$10$ and $J$=$1$.
(c) and (d) denote the charge gap for density $n$=$5$ and $L$=$12$.
(c) contains $\Delta$ vs. $U'$ at $J$=1, while (d) shows
$\Delta$ as a function of $J$ at $U'$=$11$.
}
\end{figure}

To investigate the charge excitations,
it is useful to measure the charge gap, defined as
$\Delta$=$E(N_{e}+2)+E(N_{e}-2)-2E(N_{e})$,
where $E(N_{e})$ denotes the lowest energy in the subspace
with the total number of electrons $N_{e}$.
In Fig.~4(a), the charge gap is shown as a function of $1/L$
at densities $n$=$4$ and $5$,
for particular values of $U_{\rm eff}$ and $J$.
Clearly, at these densities the charge gap extrapolates
to a nonzero value in the thermodynamic limit,
indicating that the system is an $insulator$.
On the other hand, as shown in Fig.~4(b),
for non-integer electron densities,
the charge gaps seem to extrapolate to zero
in the thermodynamic limit,
suggesting a metallic behavior.
These results indicate that a transition from an insulating phase
to a metallic regime is obtained
by changing the density away from $n$=5.

In Fig.~4(c), the charge gap for the density $n$=$5$ and $L$=$12$
is presented. It appears that $U'$ is the main driver of the system
into an insulating phase.
On the order hand, the Hund's coupling $J$ has the opposite effect:
As observed in Fig.~4(d), by increasing $J$ the charge gap decreases.
Note that $U'$ plays a role similar to that of the nearest-neighbor
Coulomb repulsion $V$ in the two-leg ladder extended Hubbard model
(with the two legs playing the role of the two orbitals in our model).
In the ladder case, it is known that $V$ drives the system to
an insulator at quarter-filling.\cite{daulnoack}

\begin{figure}[t]
\begin{center}
\includegraphics[width=0.8\linewidth]{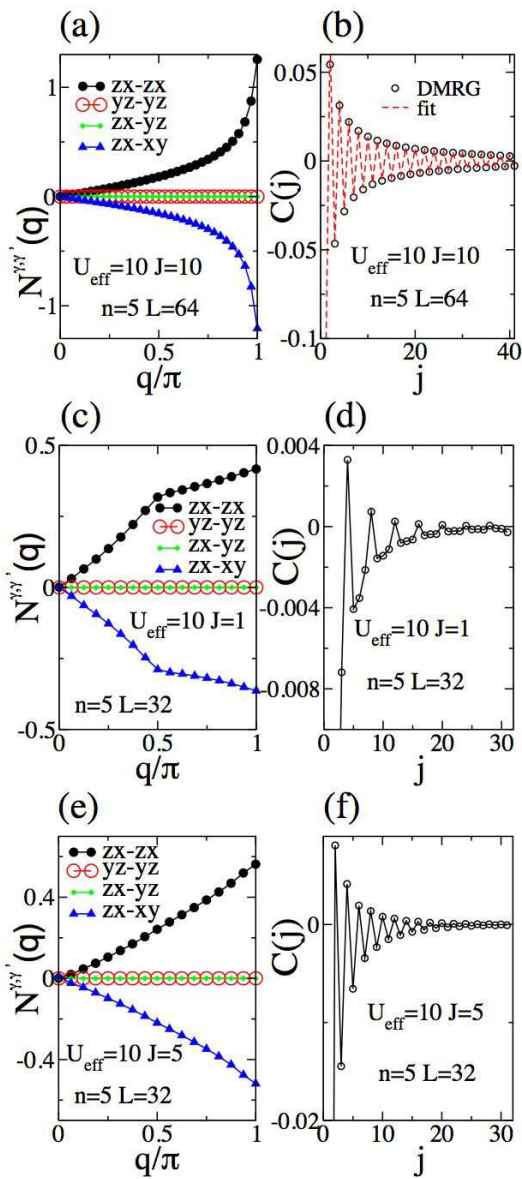}
\end{center}
\caption{\label{fig5}
The charge structure factor $N^{\gamma,\gamma'}(q)$
and the charge-charge correlation function $C(j)$, at density $n$=$5$.
(a) and (b) are for $U_{\rm eff}$=$10$, $J$=$10$ and $L$=$64$. 
This corresponds to the FM regime of Fig.~1.
The dashed line is a fit using the function $a\cos(\pi j)/j$
with an appropriate fitting parameter $a$.
(c) and (d) are for $U_{\rm eff}$=$10$, $J$=$1$, and $L$=$32$.
This is in the PM regime of Fig.1.
(e) and (f) are the same as (c) and (d), respectively,
but for $J$=$5$.
}
\end{figure}

We have also investigated the charge structure factor, defined as
\begin{equation}
 N^{\gamma,\gamma'}(q)=
 \frac{1}{2L} \sum_{j,k}e^{iq(j-k)}
 \left(N^{\gamma,\gamma'}(j,k)+N^{\gamma',\gamma}(j,k)\right),
\end{equation}
where $N^{\gamma,\gamma'}(j,k)$=%
$\langle \delta n_{\gamma}(j)\delta n_{\gamma'}(k)\rangle$ and
$\delta n_{\gamma}(j)$=$n_{\gamma}(j)-\langle n_{\gamma}(j)\rangle$.
In a periodic system $N^{\gamma,\gamma'}(j,k)$=$N^{\gamma',\gamma}(j,k)$.
However, with open boundary conditions, as used in our investigation,
this is not valid any more, due to the presence of Friedel oscillations.
Using the definition discussed above, $N^{\gamma,\gamma'}(q=0)$
is always \emph{zero}.
In our calculations, we obtained $N^{\gamma,\gamma'}(q=0)$$<$$10^{-4}$,
indicating that we have retained enough states in the truncation process
to satisfy this constraint.

\begin{figure}[t]
\includegraphics[width=0.8\linewidth]{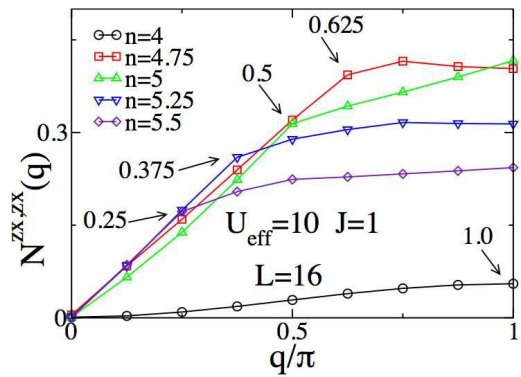}
\caption{\label{fig6}
The charge structure factor $N^{zx,zx}(q)$ for several densities
and using $U_{\rm eff}$=$10$, $J$=$1$, and $L$=$16$.
The arrows indicate the cusp positions.
}
\end{figure}

The best indication of a true long-range-order (LRO) can be obtained
by the system-size dependence of $N^{\gamma,\gamma'}(q)$.
If $N^{\gamma,\gamma'}(q^{*})/L\rightarrow constant$
as $L\rightarrow\infty$, at some particular $q^{*}$,
a true LRO characterized by $q^{*}$ is present.
Carrying out this analysis, we have found no evidence of LRO
in the charge sector of $n$=5.
In Figs.~5(a), (c), and (e), typical examples of the charge structure
factor for the FM and PM phases at density $n$=$5$ are presented.
In the FM phase, we are able to explore very large system sizes,
since we can measure the correlations in the sector of
$S_{\rm total}^{z}$=max,
with a much smaller Hilbert space than for the PM phase.
Although we did not find LRO, the behavior of the structure factor
suggests that in the FM phase the charge-charge correlation presents
a quasi-LRO due to the presence of a robust peak at $q$=$\pi$.
In fact, in the charge-charge correlation function,
defined as
\begin{equation}
  C(l)=\frac{1}{M}\sum_{|i-j|=l}
  \left\langle \delta n_{zx}(i)\delta n_{zx}(j)\right\rangle,
\end{equation}
we observe a slow power-law decay, as shown in Fig.~5(b).
This correlation oscillates as $\cos(\pi j)/j$,
as indicated by the dotted curve in Fig.~5(b).
The DMRG data agree very nicely with a fit using this function.
These strong charge oscillations suggest that the system may
develop LRO rapidly, when a coupling in the direction perpendicular
to the chains is introduced.
Then, spin-FM charge-ordered states should be seriously considered
as a possibility for Co-oxide materials, although more detailed
calculations are needed to confirm this speculation.

Note also that the negative values of $N^{zx,xy}(q=\pi)$
suggest an alternation of charge occupation
between the $zx$ and $xy$ orbitals,
as in the schematic representation in Fig.~1 (FM phase).
Indeed, as discussed later in more detail,
there is quasi-long-range AFO order.
A similar result has already been found in the FM phase
for the density $n$=$4$.\cite{onishihotta}

On the other hand, in the PM phase,
$N^{\gamma,\gamma'}\left(q\right)$ does not present a peak
as sharp as for the FM phase, as shown in Fig.~5(c).
In fact, the magnitude of the charge correlations is drastically
different between the PM and FM phases, as can be seen
from the absolute values of these correlations
in the vertical axes of Figs.~5(b) and (d).
Also note that the appearance of the cusp at $q$=$\pi/2$
is related to the four-site periodicity
of the correlation $C(l)$, as shown in Fig.~5(d).

Our results also suggest that the charges behave differently
in two distinct regimes in the PM phase.
At small $J$, the correlation $C(l)$ presents a four-sites periodicity,
while for larger $J$, only a two-site periodicity is found,
as observed in Fig.~5(f).
In addition, the cusp of $N^{zx,zx}(q)$ present in the small-$J$ regime
disappears at larger $J$ (Fig.~5(e)), apparently continuously.
We have also observed that at small $J$,
the position of the cusp changes with the electron density
in a similar way as $S(q)$, as shown in Fig.~6.


\subsection{Orbital correlations at $n$=5}

\begin{figure}[t]
\begin{center}
\includegraphics[width=0.8\linewidth]{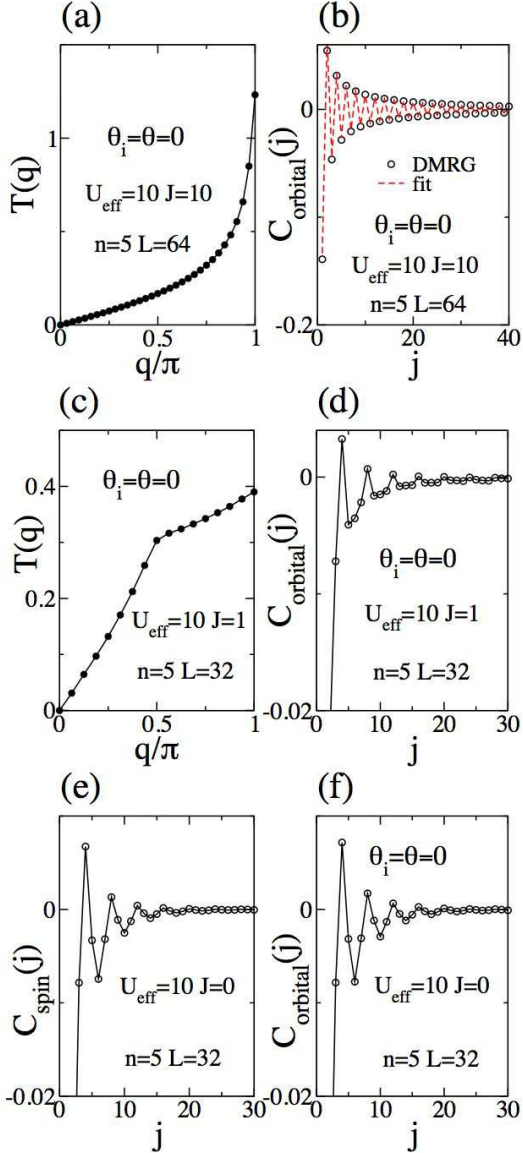}
\end{center}
\caption{\label{fig7}
(a) The orbital structure factor $T(q)$ versus momentum
for $U_{\rm eff}$=$10$, $J$=$10$, and $L$=$64$ with $\theta_i$=$\theta$=$0$.
(b) The orbital correlation $C_{\rm orbital}(j)$
for the same parameters as used in (a).
The dashed line is a fit using the function $a\cos(\pi j)/j$.
(c) and (d) are the same as (a) and (b),
but for $U_{\rm eff}$=$10$, $J$=$1$, and $L$=$32$.
(e) and (f) contain the correlations $C_{\rm spin}(j)$ and 
$C_{\rm orbital}(j)$ for $U_{\rm eff}$=$10$, $J$=$0$, and $L$=$32$.
All the results are for the density $n$=$5$.
}
\end{figure}

\begin{figure}[t]
\includegraphics[width=0.7\linewidth]{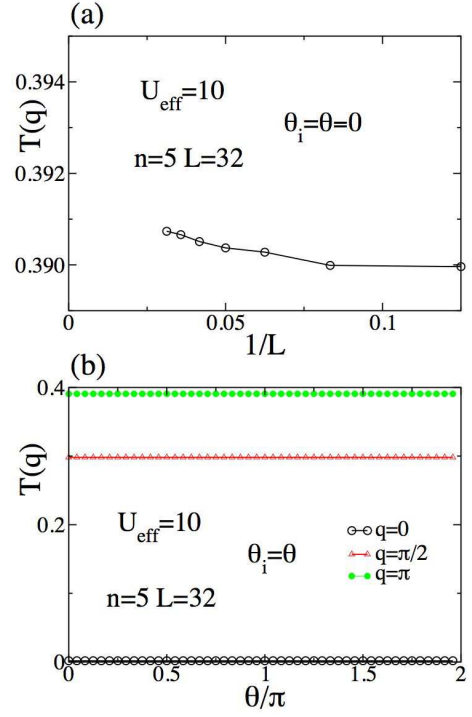}
\caption{\label{fig8}
(a) The size dependence of the orbital structure factor
$T(q)$ at $q$=$\pi$ with $\theta_{i}$=$\theta$=$0$, at density $n$=$5$.
(b) $T(q)$ vs. $\theta$ for particular values of $q$.
}
\end{figure}

Consider now the possibility of orbital order.
In the PM phase and for $n$=$5$,
we have found that the $xy$ and $zx$ orbitals are those of relevance,
since the $yz$ orbitals are fully occupied.
Note that in the PM phase and at $n$=$4$,
the orbital degree of freedom becomes inactive due to the
ferro-orbital order.\cite{onishihotta}
Then, here we take the pseudospin representation for the $xy$ and $zx$
orbitals, and measure the orbital correlations
to determine the orbital structure.
For this purpose, we introduce an angle $\theta_{j}$
to characterize the orbital shape at each site.
Using the angle $\theta_{j}$, we define the phase-dressed operators as
\begin{equation}
\left\{
\begin{array}{l}
f_{j,a,\sigma}=
 e^{i\theta_{j}/2}
  \left(\cos(\theta_{j}/2)d_{j,xy,\sigma}
       +\sin(\theta_{j}/2)d_{j,zx,\sigma}\right),\\
f_{j,b,\sigma}=
 e^{i\theta_{j}/2}
 \left(-\sin(\theta_{j}/2)d_{j,xy,\sigma}
       +\cos(\theta_{j}/2)d_{j,zx,\sigma}\right).
\end{array}
\right.
\end{equation}
The optimal orbitals, $a$ and $b$, are determined so as to maximize
the orbital structure factor, defined by
\begin{equation}
 T\left(q\right)=
 \frac{1}{L}\sum_{j,k}e^{iq(j-k)}\langle T^{z}(i)T^{z}(j) \rangle,
\end{equation}
where $T^{z}(j)$=%
$\sum_{\sigma}(f_{j,a,\sigma}^{\dagger}f_{j,a,\sigma}
-f_{j,b,\sigma}^{\dagger}f_{j,b,\sigma})/2$.

Let us first focus on the case $\theta_{i}$=$\theta$=0.
In Figs.~7(a) and (c), typical examples of the orbital structure factor
in the FM and PM phases at density $n$=$5$ are presented. 
Note that these results are similar to those of the charge structure factor
shown in Figs.~5(a) and (c), as previously anticipated.
Also, as shown in Figs.~7(b) and (d),
concerning the orbital correlation function defined as
\begin{equation}
  C_{\rm orbital}(l)
  =\frac{1}{M}\sum_{|i-j|=l}\langle T^{z}(i)T^{z}(j) \rangle,
\end{equation}
we find the same form as $C(l)$,
as observed in Figs.~5(b) and (d).
In the FM phase, as shown in Fig.~7(b),
$C_{\rm orbital}(l)$ decays as $\cos(\pi j)/j$,
which is the signature of quasi-long-range AFO.
On the other hand, in the PM phase,
we observe a four-site periodicity of $C_{\rm orbital}(j)$
as well as that of $C_{\rm spin}(j)$,
while the peak position of $T(q)$ is at $q$=$\pi$
for $U_{\rm eff}$=10 and $J$=1.
Note that the spin-spin correlation function shows
the four-site periodicity and
$S(q)$ has the peak at $q$=$\pi/2$ for $U_{\rm eff}$=10 and $J$=1,
as shown in Fig.~3(a).

To clarify the similarity between the two-orbital model
composed of the $xy$ and $zx$ orbitals and the $SU(4)$ spin-orbital model,
we investigate $C_{\rm spin}(l)$ and $C_{\rm orbital}(l)$ 
for the present $t_{\rm 2g}$ model, at $U_{\rm eff}$=$10$ and $J$=0.
As shown in Figs.~7(e) and (f),
it is clearly observed that $C_{\rm orbital}(l)$ and $C_{\rm spin}(l)$
present exactly the same behavior with a four-site periodicity,
due to the presence of the $SU(4)$ symmetry at $J$=0.
When we include the effect of $J$, the spin and orbital degrees of freedom
are no longer equivalent, but we can observe the four-site periodicity
in both $C_{\rm orbital}(l)$ and $C_{\rm spin}(l)$
due to the influence of the $SU(4)$ symmetry at $J$=0.
Thus, the short-range orbital correlation for small $J$ originates
in the $SU(4)$ singlet at $J$=0.

It should be mentioned that there is no indication of orbital LRO
in the PM phase, since $T(\pi)$ converges to a finite value
in the thermodynamic limit, as shown in Fig.~8(a).
On the other hand, although we have found no signature of orbital order
between the $xy$ and $zx$ orbitals through the orbital structure factor
$T(q)$ for $\theta_{i}$=$\theta$=0,
a more complex combination between these orbitals could exist,
but we cannot observe it directly from $T(q)$ with $\theta$=0.
In order to consider other combinations,
we set $\theta_{i}$=$\theta$ and change the value of $\theta$.
However, even in this case, we do not observe any changes in $T(q)$,
as observed in Fig.~8(b).
Namely, the orbital correlation does not change due to the rotation
in the orbital space, and we cannot determine the optimal orbitals.
Note that even if we optimize $\theta_{i}$ at each site,
$T(\pi)$ is always maximum.
Thus, we conclude that the states considered in our investigations
do not have long-range orbital order in the PM phase.


\section{Conclusions}

In this paper, we have investigated the properties of the one-dimensional
Hubbard model with three active orbitals,
with emphasis on electron densities of relevance for cobalt oxides.
We envision this work as a first step toward a numerical accurate study of
many-body Hamiltonians for Co-oxides including the Coulombic repulsion.
Our main result is the identification of two dominant tendencies
in the ground state.
For example, at sufficiently large Hund's coupling,
a tendency toward a fully saturated ferromagnetic state exists.
This state may develop long-range charge order in the case
when the interorbital repulsion $U'$ is large, at density $n$=5,
and for nonzero values of the coupling
in the direction perpendicular to the chains.

In previous investigations of Co-oxides models,
tendencies toward FM were also discussed
(see, for instance, Refs.~\onlinecite{maekawa} and \onlinecite{ogata}
and references therein).
Thus, evidence is accumulating that magnetic states should be
of relevance for these materials.
This is clearly compatible with experiments for
perovskites cobaltites.\cite{cobalt-PS}
However, thus far only for large Na doping,
magnetism has been observed experimentally
in triangular lattices.\cite{cobalt-cava}
This result may arise from the competition with the higher dimensional
version of the PM state discussed in this work.
This state has short-range correlations in all channels,
and in some limits it has an extra $SU(4)$ symmetry
as in two orbital models.
While this exact symmetry may appear
only in one dimension and for $J$=0,
remnants may remain under more realistic conditions.

To the extent that our results can be qualitatively extended
to higher dimensions, the main competition in Co-oxide models
originates from FM and PM states, with long-range and short-range
spin and charge order, respectively.
Of course, the effect of geometrical frustration could be also
an important ingredient to bring about
the complex spin-charge-orbital structure
in the triangular-lattice systems.
In fact, two of the authors have revealed that the spin frustration
is suppressed due to the orbital ordering in the $e_{\rm g}$-orbital
model on a zigzag chain.\cite{Onishi-Hotta}
In addition, in the present work we have identified metal-insulator transitions
with doping away from $n$=5, while the main properties
in the spin and charge sectors remain similar
as for the integer density $n$=5.
The next challenge is to increase the dimensionality of the
$t_{\rm 2g}$ system toward two dimensions by studying ladders
and/or zigzag chains. Work is in progress in this direction.

\acknowledgements

This work was supported by DMR-0454504 (E. D. and J. C. X.)
and FAPESP-04/09689-2 (J. C. X.).
T. H. is supported by the Japan Society for the Promotion of Science
and by the Ministry of Education, Culture, Sports, Science,
and Technology of Japan.



\begin{thebibliography}{99}
\bibitem{comment-3/2}
Actually, it is expected that this exponent $3/2$
will depend on $U$, $J$ and the density $n$.
Here, we do not intend to present a systematic study
of this exponent.
Our intention is only to show that at large $U$,
it has the same exponent as the $SU(4)$ model.
Note also the similarity with the 1D Hubbard model,
where this exponent for $U\rightarrow\infty$
is also $3/2$ (see, for example, 
J. Voit, Rep. Prog. Phys. \textbf{58}, 977 (1995)).
\bibitem{comment-gap}
Note that we tried to calculate directly the finite spin-gap anticipated
from the $n$=3 and 4 results. However, this is a difficult task,
since even for small systems the spin gap is already very small
(less than $10^{-2}$).
\end{thebibliography}
\end{document}